\documentclass[iop]{emulateapj}
\usepackage{graphicx}
\usepackage{graphics}
\newcommand{\beq}{\begin{eqnarray}}
\newcommand{\eeq}{\end{eqnarray}}
\newcommand{\bdm}{\begin{displaymath}}
\newcommand{\edm}{\end{displaymath}}
\newcommand{\be}{\begin{equation}}
\newcommand{\ee}{\end{equation}}
\newcommand{\eqd}{\,\, .}
\newcommand{\eqc}{\,\, ,}
\newcommand{\pd}[1]{\, \partial #1 \,}
\newcommand{\td}[1]{\, \mathrm{d} #1 \,}
\newcommand{\intl}{\int\limits}
\newcommand{\HF}[1]{\; H\left[ #1 \right]}  
\newcommand{\DF}[1]{\; \delta\left( #1 \right)}  
\newcommand{\eps}{\epsilon}
\newcommand{\epse}{\frac{\epsilon}{E_0}}
\newcommand{\epseb}[1]{\left( \frac{\epsilon}{E_0} \right)^{#1}}
\newcommand{\ts}{t_{syn}}
\received{}
\revised{}
\accepted{}
\shorttitle{Synchrotron Lightcurves of SSC cooled electrons}
\shortauthors{M. Zacharias \& R. Schlickeiser}
\begin{document}
\title{Synchrotron Lightcurves of blazars in a time-dependent synchrotron-self Compton cooling scenario}
\author{Michael Zacharias \& Reinhard Schlickeiser}
\affil{Institut f\"ur Theoretische Physik, Lehrstuhl IV: Weltraum- und Astrophysik, Ruhr-Universit\"at Bochum, 
44780 Bochum, Germany}
\email{mz@tp4.rub.de, rsch@tp4.rub.de}
\begin{abstract}
Blazars emit non-thermal radiation in all frequency bands from radio to $\gamma$-rays. Additionally, they {often} exhibit rapid flaring events {at all frequencies} { with} doubling time scale of the TeV { and X-ray} flux on the order of minutes, and such rapid flaring events are hard to explain theoretically. { We explore} the effect of the synchrotron-self Compton cooling, which is inherently time-dependent, leading to a rapid cooling of the electrons. Having discussed intensively the resulting effects of this cooling scenario on the spectral energy distribution of blazars in previous papers, the effects of the time-dependent approach on the synchrotron lightcurve {are investigated here}. Taking into account the retardation due to the finite size of the source {and the source geometry}, we show that the time-dependent {synchrotron-self Compton (SSC) cooling} still has profound effects on the lightcurve compared to the usual {linear (synchrotron and external Compton) cooling terms. This is most obvious if the  SSC cooling takes longer than the light crossing time scale. Then in most frequency bands the variability time scale is up to an order of magnitude shorter than under linear cooling conditions.} This is yet another strong indication that the time-dependent approach should be taken into account for modeling blazar flares { from compact emission regions}. 
\end{abstract}
%
%
%
%
\section{Introduction}

Blazars, a subclass of active galactic nuclei in the accepted unification scheme of Urry \& Padovani (1995), are characterized by a broad non-thermal spectrum exhibiting two characteristic humps and stretching from radio to $\gamma$-ray frequencies. In leptonic models the low-energy component is attributed to synchrotron radiation of highly relativistic electrons, while the high-energy component is inverse Compton emission of the same electron population (for recent reviews see B\"ottcher 2007, 2012). Several { target photon fields are relevant for the inverse Compton process}. 

Jones et al. (1974) proposed the synchrotron radiation emitted by the relativistic electrons as { the target} photon field, which is then up-scattered by the same electrons, the so-called synchrotron-self Compton (SSC) process. 

The vicinity of an active galactic nucleus harbors also additional strong external (to the jet) photon fields, which can potentially contribute in the form of so-called external Compton radiation to the high-energy component of blazars. Such external fields could come from the accretion disk (Dermer \& Schlickeiser 1993), the broad line region (Sikora et al. 1994) or the dusty torus (Blazejowski et al. 2000, Arbeiter et al. 2002). These external fields are usually preferred over the SSC, if the high-energy component dominates the synchrotron component in the spectral energy distribution (SED) of blazars.

It is well established that blazars are far from being steady sources. { They} exhibit strong flares in all frequency bands, which can in some cases outshine even the brightest galactic sources. The brightest $\gamma$-ray flare ever detected is from 3C 454.3, reported by Vercellone et al. (2011), reaching a $\gamma$-ray flux of $F_{\gamma}=(6.8\pm 1.0)\times 10^{-5}$ photons cm$^{-2}$ s$^{-1}$, which is six times higher than the Vela pulsar. Additionally, blazars also exhibit very rapid flares with doubling time scales on the order of minutes as in the case of PKS 2155-304 (Aharonian et al. 2007) or PKS 1222+216 (Tavecchio et al. 2011) { in the TeV regime, or Mrk 421 in the X-rays (Cui 2004)}. 

Such rapid flares are theoretically challenging, since typical cooling time scales of the radiating electrons are considerably longer. Several models have thus been invoked to explain these rapid flares, such as the jet-in-a-jet model (Giannios et al. 2009), the similar minijets-in-a-jet model (Biteau \& Giebels 2012, Giannios 2013), magneto-centrifugal acceleration of beams of particles (Ghisellini et al. 2009a), a star traversing the jet (Barkov et al. 2012), and others. Quite common in all these models is the assumption of an emission blob being smaller than the jet { cross-section} and { moving} much faster than the surrounding { relativistic} jet material. { This gives} rise to a very short light-crossing time scale, which is usually equaled to the variability time scale.

In many theoretical investigations, as the ones cited above, and in most modeling attempts (e.g. Ghisellini et al. 2009b) the electron distribution is assumed to be { stationary}. This eases the computational effort, of course, and might be suitable for steady sources or those varying over a long time scale. { However,} it is certainly not { justified} for rapid flares as in PKS 2155-304 or PKS 1222+216. { The time-dependence of the relativistic electron distribution function has important effects on the} resulting SED, as is demonstrated in a recent series of papers (Schlickeiser 2009; Schlickeiser et al. 2010 (hereafter SBM); Zacharias \& Schlickeiser 2010, 2012a { (hereafter ZSa)}, 2012b { (hereafter ZSb)}).

Relativistic electrons in a { relativistically moving emission blob along the jet of the active galactic nucleus lose energy by emitting synchrotron radiation.} These synchrotron photons are a prime target for the same electrons to inverse Compton scatter them to higher energies. This is the SSC process, as mentioned above, { which is an additional} energy loss process for the electrons. This in turn implies that the subsequently emitted synchrotron photons are less energetic, and so will be the SSC photons. { Thus, this results} in a decreased efficiency of the SSC process and in a decreased efficiency of the { SSC energy loss process with respect to time. Consequently,} even if the SSC process dominates initially the electron losses, eventually the { time-independent} loss processes such as synchrotron and external Compton losses dominate the loss rate. Schlickeiser (2009), as well as Zacharias \& Schlickeiser (2010) were able to show that the time-dependent treatment of the SSC losses leads to a much faster electron cooling compared to the steady-state approach. 

{ Therefore, it is interesting} to discuss the effects of this rapid cooling on blazar lightcurves, where the variability can be displayed in an obvious way.

It is the purpose of this paper to highlight the different effects of the linear and the time-dependent { (nonlinear)} SSC cooling on the synchrotron lightcurves. To keep the problem simple { and analytically tractable, we} utilize only the retardation effect due to the finite size of the emission region, and the geometry of the source. This will be discussed in section \ref{sec:geo}, where we will derive the necessary formula to calculate the lightcurve from the synchrotron intensity. The latter was already calculated by SBM, and we will summarize their results in section \ref{sec:sin} for the sake of completeness. We will then use the derived formula from section \ref{sec:geo} to calculate the resulting lightcurves in sections \ref{sec:la01} and \ref{sec:la10}. We will discuss the results in section \ref{sec:dis} and conclude in section \ref{sec:con}.

{The more involved calculations of the inverse Compton lightcurves will be discussed in a future publication.}

%
%

\section{Geometry of the situation} \label{sec:geo}
\begin{figure}[tb]
	\centering
		\includegraphics[width=0.48\textwidth]{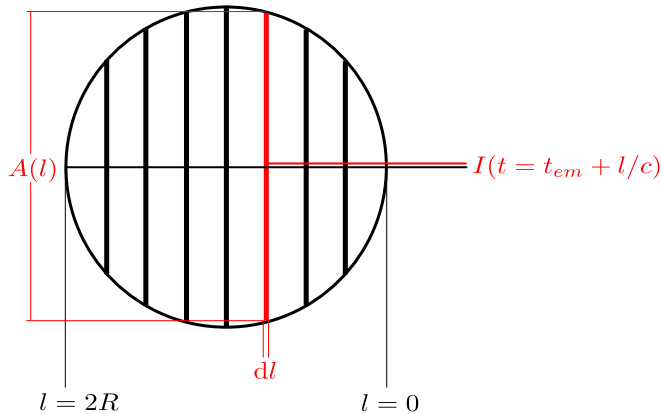}
	\caption{Sketch of the situation: The light of the slice at position $l$ (with volume $\td{V(l)} = A(l)\td{l}$) is received by the observer at time $t=t_{em}+l/c$.}
	\label{fig:geo}
\end{figure}
We assume a spherical, uniform radiation zone in the jet as depicted in figure \ref{fig:geo}.

{ For negligible} retardation the received monochromatic intensity at intrinsic time $t_{em}$ is $I(t_{em},\eps)$, where $\eps$ is the intrinsic energy of the photon. Since, however, the source has a finite size, photons emitted at the back of the source will arrive at the observer at a later time $\Delta t=2R/c$ than the photons emitted at the front, with $R$ being the radius of the spherical source and $c = 3\cdot 10^{10} \;\hbox{cm}/\hbox{s}$ the speed of light. 

{ We include this retardation effect, but assume that the source is (i) spatially homogeneous, and (ii) optically thin. For optically thin sources all photons can leave the emission region without further spatial diffusion (Eichmann et al. 2010). Then} the received intensity is just a function of the distance $l$ of the production site from the front. { Using a similar approach as Chiaberge \& Ghisellini (1999), we cut} the source into slices of length $\td{l}$, as shown in figure \ref{fig:geo}. The received intensity of each slice is
\beq
\td{I(t-l/c,l,\eps)} = I(t-l/c,\eps) \frac{\td{V(l)}}{V} \HF{t-l/c} \eqd
\label{eq:dI}
\eeq
Here, $t$ is the time of the observer, which equals $t_{em}$ for $l=0$ (the front of the source). The Heaviside function $\HF{x}$ displays the fact that light from a specific slice can only be detected after it has crossed the distance $l$ to the front of the source. 

The fraction $\td{V(l)}/V$ is a geometrical weight function, which is defined in such a way that the integral over $\td{V(l)}/V$ equals unity. Since in a spherical source each slice has a different volume than the other slices, its contribution depends on its position in the source. The volume of the slice is given by $\td{V(l)}=A(l)\td{l}$, where $A(l)=\pi(2Rl-l^2)$ is the cross section of the slice at position $l$. The geometrical weight function then becomes
\be
\frac{\td{V(l)}}{V} = \frac{\pi(2Rl-l^2)}{\frac{4}{3}\pi R^3}\td{l} = \frac{3}{R}\left[ \frac{l}{2R} - \left( \frac{l}{2R} \right)^2 \right] \td{l} \eqd
\label{eq:gwf}
\ee
The complete received monochromatic lightcurve $L(t,\eps)$ then equals the sum over the contribution { from} all slices:
\beq
L(t,\eps) = \int \td{I(t-l/c,l,\eps)} \nonumber \\
= \intl_{0}^{2R} I(t-l/c,\eps) \HF{t-l/c} \frac{3}{R}\left[ \frac{l}{2R} - \left( \frac{l}{2R} \right)^2 \right] \td{l} \nonumber \\
= 6 \intl_0^1 I(t-\lambda_0\lambda,\eps) (\lambda-\lambda^2) \HF{t-\lambda_0 \lambda} \td{\lambda} \eqc
\label{eq:L}
\eeq
after an obvious substitution. Here we { introduced} the light-crossing time scale $\lambda_0 = 2R/c$.

We note that equation (\ref{eq:dI}) is general as long as the assumptions (i) and (ii) are satisfied. Thus, it is not limited to spherical geometries, and for example cylindrical sources could also be { chosen}. In fact, a cylindrical geometry would lead to a simpler form of the geometrical weight function. However, the assumption of isotropy for the electron distribution and the radiation fields (see below) would not be valid any more.\footnote{{ For example, Chiaberge \& Ghisellini (1999) chose a cubed geometry.}}

%
%

\section{Synchrotron Intensity} \label{sec:sin}

{ In this section we summarize results previously obtained (SBM, ZSa, ZSb) in order to introduce the relevant functions and parameters.}

The isotropic, optically thin synchrotron intensity from relativistic electrons with the volume-averaged differential density $n(\gamma,t)$ is given by
\be
I_{syn}(\eps,t) = \frac{R}{4\pi}\intl_0^{\infty} n(\gamma,t)P_{syn}(\eps,\gamma) \td{\gamma} \eqc
\label{eq:monoint}
\ee
with
\be
P_{syn}(\eps,\gamma) = \frac{P_0\eps}{\gamma^2} CS\left( \frac{2\eps}{3\eps_0\gamma^2} \right)
\label{eq:Psyn}
\ee
being the synchrotron power of a single electron in a large-scale random magnetic field of constant strength $B=b\;\hbox{Gauss}$ (Crusius \& Schlickeiser 1988). Here $P_0=2\cdot 10^{24}\;\hbox{erg}^{-1}\hbox{s}^{-1}$, and $\eps_0=1.9\cdot 10^{-20}b\;\hbox{erg}$. The function $CS(x)$ is well approximated by 
\be
CS(x) \approx a_0 x^{-2/3}e^{-x} \eqc
\label{eq:cs}
\ee
with $a_0 = 1.151275$.

The differential { relativistic} electron density can be calculated from the kinetic equation (Kardashev 1962)
\be
\frac{\pd{n(\gamma,t_{em})}}{\pd{t}}-\frac{\pd}{\pd{\gamma}}\left[ |\dot{\gamma}| n(\gamma,t_{em}) \right] = S(\gamma,t_{em}) \eqc
\label{eq:kineq} 
\ee
where $|\dot{\gamma}|$ is the electron energy loss term, and $S(\gamma,t_{em})$ is the source term.

For demonstration purposes and ease of calculation we use a relatively simple source term
\be
S(\gamma,t_{em}) = q_0 \DF{\gamma-\gamma_0} \DF{t_{em}} \eqc
\label{eq:source}
\ee
that is a single injection of monochromatic electrons with the injection Lorentz factor $\gamma_0$ and the electron density $q_0$.

In the scenario depicted here we consider {electron losses via the synchrotron, external Compton} and synchrotron-self Compton channels. Since the latter depends on the produced synchrotron radiation, and thus directly on the electron distribution, the kinetic equation becomes non-linear (Schlickeiser 2009). The total electron loss term is given by
\beq
|\dot{\gamma}| = |\dot{\gamma}|_{syn} + |\dot{\gamma}|_{ec} + |\dot{\gamma}(t_{em})|_{ssc} \nonumber \\
= D_0 (1+l_{ec}) \gamma^2 + A_0\gamma^2\intl_0^{\infty} \gamma^2 n(\gamma,t_{em}) \td{\gamma} \eqd
\label{eq:elloss}
\eeq
The parameters are $D_0=1.3\cdot 10^{-9} b^2\;\hbox{s}^{-1}$, and $A_0=1.2\cdot 10^{-18}R_{15}b^2\;\hbox{cm}^3\hbox{s}^{-1}$, where we scaled the radius of the source as $R=10^{15}R_{15}\;\hbox{cm}$. 

{We define
\be
l_{ec} = \frac{|\dot{\gamma}|_{ec}}{|\dot{\gamma}|_{syn}} = \frac{4\Gamma_b^2}{3}\frac{u_{ec}^{\prime}}{u_B} \eqd
\label{eq:lecdef}
\ee
where $\Gamma_b$ is the Lorentz factor of the plasma blob, $u_{ec}^{\prime}$ is the isotropic energy density of the external radiation field in the frame of the host galaxy, and $u_B$ is the energy density of the magnetic field. This parameter describes the relative strength of external to synchrotron cooling, and has profound consequences for the SED, as we showed in Zacharias \& Schlickeiser (2012b). We note that it is { less} important for the discussion of synchrotron lightcurves and only introduced for the sake of completeness.}

{More importantly, as one can see from equation (\ref{eq:elloss}), is the fact that} the SSC cooling term { by its dependence on $n(\gamma,t)$} is time-dependent, which means that {its strength} decreases over time. Consequently, even if the SSC cooling dominates the total cooling term initially, after some time the SSC cooling will become weaker than the linear cooling, and thus the synchrotron or external Compton cooling will dominate for later times. Obviously, if the { linear cooling terms are stronger than the SSC cooling at the beginning, they} will be stronger for all times.

This can be further quantified by the injection parameter
\be
\alpha = \sqrt{\frac{A_0q_0}{D_0 (1+l_{ec})}}\gamma_0 \eqd
\label{eq:alpha}
\ee
It is defined in such a way that 
\beq
\alpha^2=\frac{|\dot{\gamma}(t_{em}=0)|_{ssc}}{|\dot{\gamma}|_{syn} + |\dot{\gamma}|_{ec}} \eqd
\label{eq:alphadef}
\eeq
As a consequence { (ZSa and ZSb)} the { Compton dominance in the SED} depends on $\alpha^2$, at least in the Thomson limit. This demonstrates the importance of this parameter, which can also be expressed as
\be
\alpha = 46 \frac{\gamma_4N_{50}^{1/2}}{R_{15} (1+l_{ec})^{1/2}} \eqc
\label{eq:alphascaled}
\ee
where we scale the total number of electrons $N=10^{50}N_{50}$, and the initial electron Lorentz factor $\gamma_0=10^{4}\gamma_4$. { Obviously,} $\alpha$ increases for increasing $\gamma_0$ and $N$, and decreases for increasing $R$ {and $l_{ec}$}. If $\alpha\gg 1$ the cooling will initially be dominated by the SSC cooling, while for $\alpha\ll 1$ the cooling is dominated by the linear terms for all times.

{ We note that both inverse Compton cooling terms operate in the Thomson limit. In the Klein-Nishina limit the efficiencies of both cooling terms are much reduced, and become unimportant compared to the synchrotron cooling. This resembles the case $\alpha\ll 1$ and $l_{ec}\ll 1$ and is, therefore, covered by our approach.}

The differential equation (\ref{eq:kineq}) with the loss term (\ref{eq:elloss}) and the source term (\ref{eq:source}) has been solved by SBM. For $\alpha\ll 1$ (i.e. negligible SSC-losses) they obtained
\be
n(\gamma,t_{em}) = q_0 \DF{\gamma-\frac{\gamma_0}{1+D_0(1+l_{ec})\gamma_0t_{em}}} \eqc
\label{eq:na01}
\ee
which is, indeed, a linear cooling solution. \\
For $\alpha\gg 1$ (i.e. initially dominating SSC-losses) SBM found
\beq
n(\gamma,t_{em}<t_c) = q_0 \HF{t_c-t_{em}} \nonumber \\
\times \DF{\gamma-\frac{\gamma_0}{(1+3\alpha^2D_0(1+l_{ec})\gamma_0t_{em})^{1/3}}} \eqc
\label{eq:na10a}
\eeq
yielding a nonlinear dependence of $\gamma$ on time. For later times the electron density { approaches}
\beq
n(\gamma,t_{em}>t_c) = q_0 \HF{t_{em}-t_c} \nonumber \\
\times \DF{\gamma-\frac{\gamma_0}{\frac{1+2\alpha^3}{3\alpha^2}+D_0(1+l_{ec})\gamma_0t_{em}}} \eqc
\label{eq:na10b}
\eeq
which is a modified linear cooling solution. The transition time is defined as
\be
t_c = \frac{\alpha^3-1}{3\alpha^2D_0(1+l_{ec})\gamma_0} \eqd
\label{eq:tc}
\ee

The intensity (\ref{eq:monoint}) for both cases of $\alpha$ has also been calculated by SBM. For $\alpha\ll 1$ they obtained with equation (\ref{eq:na01})
\beq
I_{syn}(t_{em},\eps) = I_0 \left( \epse \right)^{1/3} \left( 1+ \frac{t_{em}}{\ts} \right)^{2/3} \nonumber \\
\times e^{-\epse \left( 1+\frac{t_{em}}{\ts} \right)^2} \eqd
\label{eq:Iunret0}
\eeq
For $\alpha\gg 1$ with equations (\ref{eq:na10a}) and (\ref{eq:na10b})
\beq
I_{syn}(t_{em}<t_c,\eps) = I_0 \epseb{1/3} \left( 1+ \frac{3\alpha^2}{\ts}t_{em} \right)^{2/9} \nonumber \\
\times e^{-\epse \left( 1+\frac{3\alpha^2}{\ts}t_{em} \right)^{2/3}} \eqc
\label{eq:Iunret1}
\eeq
and
\beq
I_{syn}(t_{em}>t_c,\eps) = I_0 \epseb{1/3} \left( \alpha_g+ \frac{t_{em}}{\ts} \right)^{2/3} \nonumber \\
\times e^{-\epse \left( \alpha_g+\frac{t_{em}}{\ts} \right)^2} \eqd
\label{eq:Iunret2}
\eeq
Here we used the definitions $I_0=3a_0RP_0q_0\eps_0/(8\pi)$, $\ts=1/(D_0(1+l_{ec})\gamma_0)$, $E_0=3\eps_0\gamma_0^2/2$, and $\alpha_g=(1+2\alpha^3)/(3\alpha^2)$.

These intensities are equal to a monochromatic lightcurve, where the retardation and, thus, the source's finite size have not been taken into account. Below, we will refer to them as the ``unretarded'' lightcurves.

Now, we have collected all necessary ingredients to calculate the retarded synchrotron lightcurves, which we present in the following sections.

%
%

\section{Monochromatic synchrotron lightcurve for dominating linear cooling} \label{sec:la01}

Using equation (\ref{eq:Iunret0}) in equation (\ref{eq:L}) we obtain the retarded lightcurve for the case $\alpha\ll 1$:
\beq
L(t,\eps) = 6I_0 \epseb{1/3} \intl_0^1 \left( 1+\frac{t-\lambda_0\lambda}{\ts} \right)^{2/3} \nonumber \\
\times e^{-\epse \left( 1+\frac{t-\lambda_0\lambda}{\ts} \right)^{2}} \left( \lambda-\lambda^2 \right) \HF{t-\lambda_0 \lambda} \td{\lambda} \eqd
\label{eq:I0}
\eeq

The integral (\ref{eq:I0}) can be solved in terms of several incomplete Gamma-functions. However, this would not give many insights. Instead, we will use meaningful approximations for the integral in three time domains. These domains can later be glued together to give a continuous analytic result. 

{ First of all, we define two characteristic} time scales of the unretarded lightcurve. They can later be connected to the light-crossing time scale, yielding some information about the resulting retarded lightcurve. The first one is the local maximum of the unretarded lightcurve, which is:
\be
t_1(\eps) = \ts \left( \sqrt{\frac{E_0}{3\eps}}-1 \right) \eqd
\label{eq:t10}
\ee
{ This expression is negative for $\eps>E_0/3$, indicating that} for such energies there is no local maximum. If $t_1(\eps)>\lambda_0$ the variability will mostly take place for times later than the light-crossing time scale. Solving the resulting inequality for $\eps$, results in
\be
\eps < \eps_1 = \frac{E_0}{3\left( 1+\frac{\lambda_0}{\ts} \right)^2} < \frac{E_0}{3} \eqd
\label{eq:e10}
\ee
This equation implies that for energies $\eps<\eps_1$ the variability due to the flare will be longer than the light-crossing time scale. Hence, we expect the global maximum of the lightcurve to occur later than $\lambda_0$, and thus be unaffected by the retardation. 

{ The second characteristic time scale is related to the argument of the exponential in the unretarded lightcurve $A=\epse\left( 1+t_{em}/\ts \right)^{2}$. As soon as $t_{em}\geq \ts$ the unretarded lightcurve exponentially decays, which should also be visible in the retarded lightcurve. Since, however, $A\approx \epse$ for $t_{em}\ll \ts$, we set
\beq
A &=& \epse \left( 1+\frac{t_{em}}{\ts} \right)^{2} \nonumber \\
&=& \epse + A^{*}(\eps,t_{em}) \eqc
\label{eq:t20def}
\eeq
with
\be
A^{*}(\eps,t_{em}) = \epse \left[ \left( 1+\frac{t_{em}}{\ts} \right)^2 - 1 \right] \eqd
\ee
Once $A^{*}(\eps,t_{em})$ is larger than unity the unretarded lightcurve will exponentially decay. Thus, we obtain the second characteristic time scale $t_2(\eps)$ by $A(\eps,t_2(\eps))=1$, yielding
\be
t_2(\eps) = \ts \left( \sqrt{1+\frac{E_0}{\eps}} -1 \right) \eqd
\label{eq:t20}
\ee
Unlike $t_1(\eps)$, the second characteristic time scale exhibits no restrictions by $\eps$.} Obviously, $t_1(\eps)<t_2(\eps)$. For $t_2(\eps)>\lambda_0$ the exponential will become important only after the light-crossing time scale. Solving the inequality for $\eps$ we obtain
\be
\eps < \eps_2 = \frac{E_0}{\left( 1+\frac{\lambda_0}{\ts} \right)^2 -1} \eqd
\label{eq:e20}
\ee

We can now begin with the actual calculation of the retarded lightcurve. The simplest case is obviously for $t>\lambda_0$, since in this case the retarded lightcurve should be the same as the unretarded lightcurve. This is due to the fact that the retardation is not important for time scales much longer than $\lambda_0$. Inspecting the difference $t-\lambda_0\lambda$, we see that $\lambda_0\lambda$ can be at most equal to $\lambda_0$. Thus, for $t\gg \lambda_0$ we can approximate $t-\lambda_0\lambda \approx t$. Hence,
\beq
L(t>\lambda_0,\eps) \approx 6I_0 \epseb{1/3} \left( 1+\frac{t}{\ts} \right)^{2/3} \nonumber \\ 
\times e^{-\epse\left( 1+\frac{t}{\ts} \right)^2} \intl_0^1 \left( \lambda-\lambda^2 \right) \td{\lambda} \nonumber \\
= I_0 \epseb{1/3} \left( 1+\frac{t}{\ts} \right)^{2/3} e^{-\epse\left( 1+\frac{t}{\ts} \right)^2} \eqc
\label{eq:I01}
\eeq
which, indeed, equals the unretarded lightcurve. 

The other rather simple case is for $t<\lambda_0$ with the further requirement that $t<t_{1,2}(\eps)$ (the subscript refers to both $t_1$ and $t_2$). The latter implies that the unretarded lightcurves were neither variable nor have they decayed already. { Then} in eq. (\ref{eq:I0}) the terms $(t-\lambda_0\lambda)/\ts$ can be neglected compared to unity, { yielding}
\beq
L(t<\lambda_0,\eps) \approx 6I_0 \epseb{1/3} e^{-\epse} \intl_0^{t/\lambda_0} (\lambda-\lambda^2) \td{\lambda} \nonumber \\
= 3I_0 \epseb{1/3} e^{-\epse} \left( \frac{t}{\lambda_0} \right)^2 \left[ 1-\frac{2}{3}\frac{t}{\lambda_0} \right] \eqd
\label{eq:I02}
\eeq
{ For times below the light-crossing time scale and below the variability time scale of the unretarded lightcurve the retarded lightcurve increases rapidly $L \propto t^2$. }

For intermediate times the calculation is quite involved, and the details can be found in appendix \ref{app:intpart}. { We obtain}
\beq
L(t_{1,2}<t<\lambda_0,\eps) = 6I_0 \epseb{1/3} \nonumber \\
\times \intl_0^{t/\lambda_0} \left( 1+\frac{t-\lambda_0\lambda}{\ts} \right)^{2/3} e^{-\epse \left( 1+\frac{t-\lambda_0\lambda}{\ts} \right)^{2}} \left( \lambda-\lambda^2 \right) \td{\lambda} \nonumber \\
\approx 3I_0 \epseb{-2/3} e^{-\epse} \frac{\ts^2}{\lambda_0^2} \left( \frac{t}{\ts} \right) \left[ 1-\frac{t}{\lambda_0} \right] \eqd
\label{eq:I03}
\eeq
We note that the exact form of the intermediate regime is not so important, since it will be glued to the approximation (\ref{eq:I02}) at $t\approx t_2$. The most important result is the linear { increase} of the the lightcurve (\ref{eq:I03}), which leads to a break at $t_2$ in the retarded lightcurve. However, if $t_{1,2}(\eps)>\lambda_0$ the intermediate part does not play a role, and the lightcurve breaks immediately at $t=\lambda_0$ from the { initial} $t^2$-dependence to the time dependence given by equation (\ref{eq:I01}). 

Depending on the { synchrotron photon} energy $\eps$, we can now construct the lightcurves from the three approximations (\ref{eq:I01}) - (\ref{eq:I03}). { We obtain two cases, divided in additional sub-cases}. 

Beginning with $\eps<E_0/3$, we get:
\beq
L(t,\eps<\eps_1) = 3I_0 \epseb{1/3} \frac{\left( \frac{t}{\lambda_0} \right)^2}{1+3\left( \frac{t}{\lambda_0} \right)^2} \nonumber \\
\times \left( 1+\frac{t}{\ts} \right)^{2/3} e^{-\epse \left( 1+\frac{t}{\ts} \right)^2} \label{eq:I011} \eqc
\eeq
\beq
L(t,\eps_1<\eps<\eps_2) = 3I_0 \epseb{1/3} \frac{\left( \frac{t}{\lambda_0} \right)^2}{\left( 1+\frac{t}{t_2} \right)^{5/3}} \nonumber \\
\times \left( 1+\frac{t}{\ts} \right)^{2/3} e^{-\epse \left( 1+\frac{t}{\ts} \right)^2} \label{eq:I012} \eqc 
\eeq
\beq
L(t,\eps_2<\eps<E_0/3) = 3I_0 \epseb{1/3} e^{-\epse} \frac{\left( \frac{t}{\lambda_0} \right)^2}{\left( 1+\frac{t}{2t_2} \right)^{5/3}} \nonumber \\
\times \left( 1+\frac{t}{\ts} \right)^{2/3} \left[ 1-\frac{t}{\lambda_0} \right]  \label{eq:I013} \eqd
\eeq
For $\eps>E_0/3$ the solutions become:
\beq
L(t,E_0/3<\eps<\eps_2) = 3I_0 \epseb{1/3} \frac{\left( \frac{t}{\lambda_0} \right)^2}{1+3\left( \frac{t}{\lambda_0} \right)^2} \nonumber \\
\times \left( 1+\frac{t}{\ts} \right)^{2/3} e^{-\epse \left( 1+\frac{t}{\ts} \right)^2} \label{eq:I021} \eqc
\eeq
\beq
L(t,\eps_2<\eps) = 3I_0 \epseb{1/3} e^{-\epse} \frac{\left( \frac{t}{\lambda_0} \right)^2}{1+\frac{t}{2t_2}} \left[ 1-\frac{t}{\lambda_0} \right] \label{eq:I022} \eqd
\eeq
The lightcurves (\ref{eq:I013}) and (\ref{eq:I022}) cut off at $t=\lambda_0$. Obviously, light from the back reaches the observer only at later times, causing the radiation to be visible on longer time scales than implied by the unretarded lightcurve. 

\begin{figure*}[t]
\begin{minipage}[t]{0.49\linewidth}
\centering \resizebox{\hsize}{!}
{\includegraphics{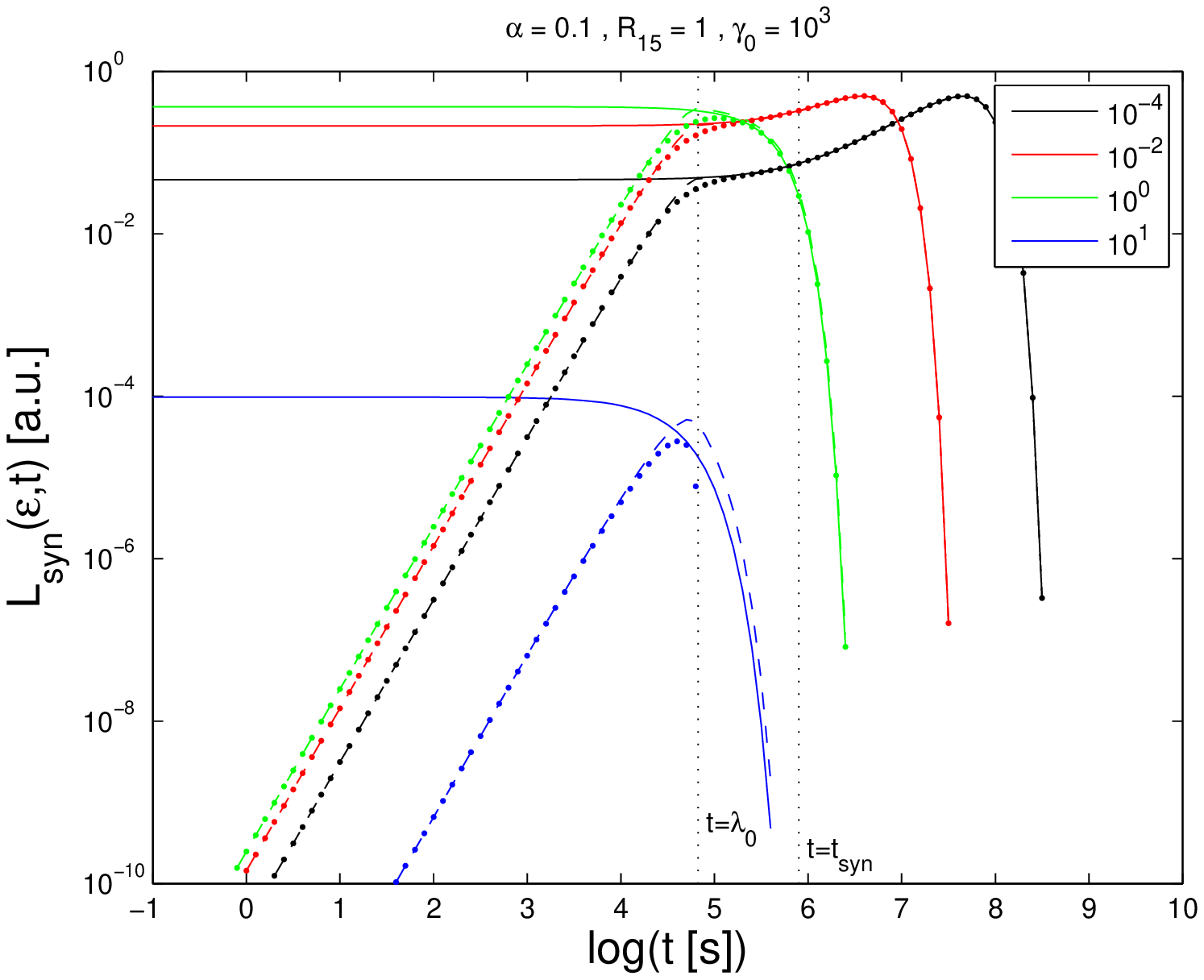}}
\end{minipage}
\hspace{\fill}
\begin{minipage}[t]{0.49\linewidth}
\centering \resizebox{\hsize}{!}
{\includegraphics{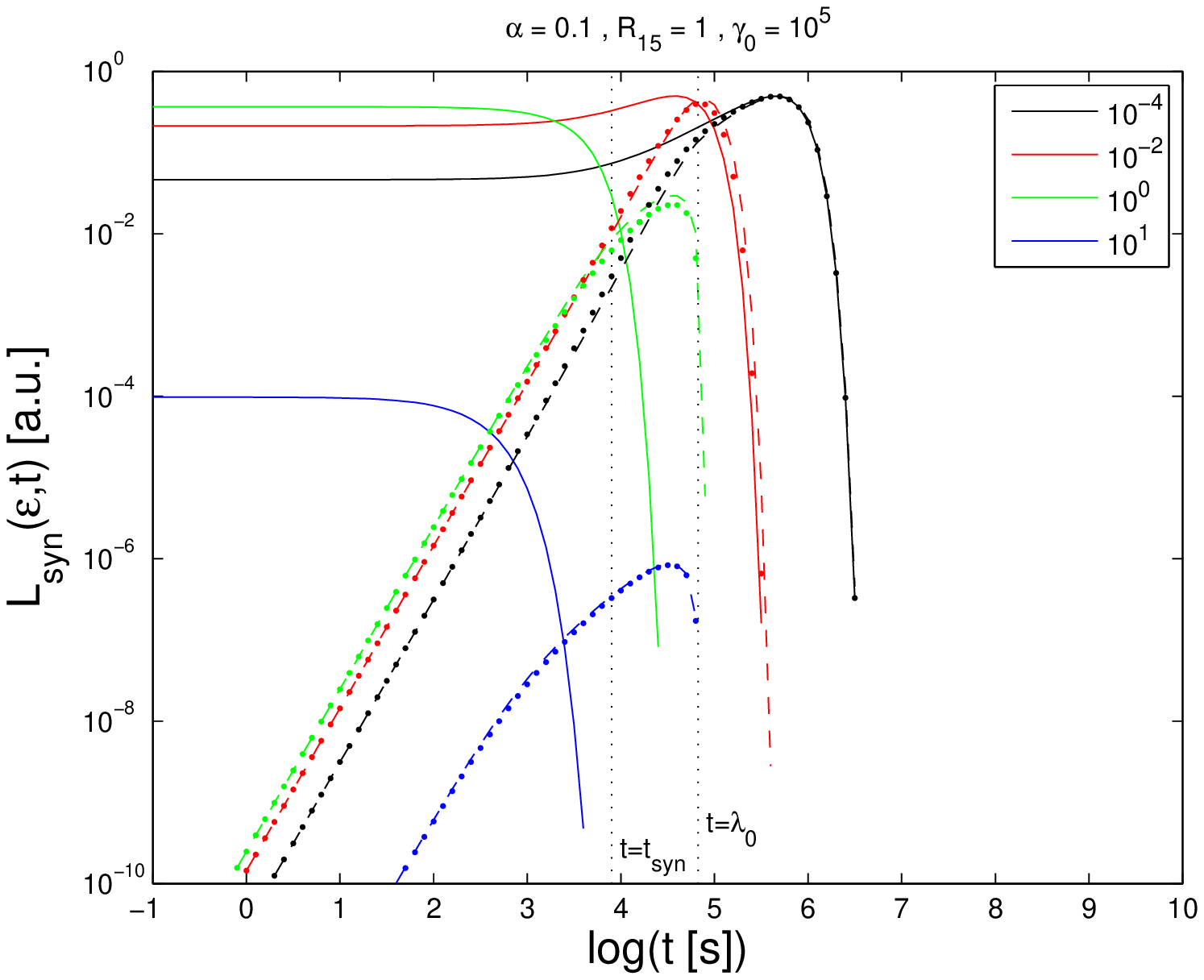}}
\end{minipage}
\caption{Unretarded (full), as well as numerical (dashed) and analytical (dotted) retarded lightcurve for $\alpha\gg 1$ and two cases of $\gamma_0$ { over a logarithmic time-axis}. The values of $\eps$ in the legend are given in units of $E_0$. The curves are normalized with $I_0$ and we set $b=1$.}
\label{fig:a01}
\end{figure*}

The analytical results (\ref{eq:I011}) - (\ref{eq:I022}) are plotted along with a numerical integration of equation (\ref{eq:I0}) in Figure \ref{fig:a01} for two cases of $\gamma_0$. For comparison, we also show the unretarded lightcurve.

The first obvious result is that the retarded synchrotron lightcurve increases rapidly as long as $t<\lambda_0$. Afterwards the retarded lightcurve behaves as the unretarded one, which is reasonable, as we discussed above. 

The other points mentioned earlier are also quite obvious. Even though the unretarded lightcurve for very high energies cuts off long before the light-crossing time scale, the retarded lightcurves are extended until $\lambda_0$. The break in the lightcurve in the intermediate time regime is also evident. However, as discussed above, the low energetic cases, where the variability time scales are much longer than the light-crossing time scale, do not exhibit this break. 

As one can see, the analytical result matches the numerical integration rather well, which is reassuring and validates a posteriori our approximations. However, there is one caveat: The distinction of cases by $t_2$ and $\eps_2$ is rather sharp (esp. equations (\ref{eq:I013}) and (\ref{eq:I022})).{ This is obvious in the left plot of Figure \ref{fig:a01} in the analytical curve for $\eps=10E_0$, which cuts off at $t=\lambda_0$. On the other hand, the numerical curve in this case decays exponentially.} The distinction of the cases divided by $\eps_2$ is, therefore, not as strict as implied by the analytical result. It is a more gradual transition, which is, however, difficult to implement in one equation. 

The problem is probably due to the rather artificial definition of $t_2$, which is also indicated by the fact that the break for the high-energy lightcurves is better placed at $2t_2$ instead of $t_2$.

%
%

\section{Monochromatic synchrotron lightcurve for dominating initial SSC cooling} \label{sec:la10}

For the case $\alpha\gg 1$ we use equations (\ref{eq:Iunret1}) and (\ref{eq:Iunret2}) in equation (\ref{eq:L}) to obtain the retarded lightcurve
\beq
L_1(t,\eps) = 6I_0 \epseb{1/3} \intl_0^1 \left( 1+\frac{3\alpha^2}{\ts}(t-\lambda_0\lambda) \right)^{2/9} \nonumber \\
\times e^{-\epse \left( 1+\frac{3\alpha^2}{\ts}(t-\lambda_0\lambda) \right)^{2/3}} \left( \lambda-\lambda^2 \right) \nonumber \\
\times \HF{t-\lambda_0 \lambda} \HF{t_c-(t-\lambda_0\lambda)} \td{\lambda} \eqc 
\label{eq:I1}
\eeq
\beq
L_2(t,\eps) = 6I_0 \epseb{1/3} \intl_0^1 \left( \alpha_g+\frac{t-\lambda_0\lambda}{\ts} \right)^{2/3} \nonumber \\
\times e^{-\epse \left( \alpha_g+\frac{t-\lambda_0\lambda}{\ts} \right)^{2}} \left( \lambda-\lambda^2 \right) \nonumber \\
\times \HF{t-\lambda_0 \lambda} \HF{(t-\lambda_0\lambda)-t_c} \td{\lambda} \eqd
\label{eq:I2}
\eeq
{ For $t_c<t<t_c+\lambda_0$ both $L_1$ and $L_2$ contribute to the emitted lightcurve, which differs from the strict division of the unretarded lightcurves (\ref{eq:Iunret1}) and (\ref{eq:Iunret2}). This is, again, an effect of the retardation: Even if light received from the front of the source is from electrons already cooling in the linear regime ($t_{em}>t_c$), the light received from the back of the source is still from electrons cooling in the nonlinear regime ($t_{em}<t_c$).} If $t_c<\lambda_0$ this period can be quite extended.

Although there are { several sub-cases to consider in the analytical calculation, we can} use the same approximation for the integrals (\ref{eq:I1}) and (\ref{eq:I2}), as we used to obtain equations (\ref{eq:I01}) - (\ref{eq:I03}). It is therefore unnecessary to repeat them in detail. Instead, we will summarize the results in the most compact form possible, where the sub-cases are combined in such a way that the resulting lightcurve is continuous. 

{ The characteristic time scales $t_3(\eps)$ and $t_4(\eps)$ are obtained by the same arguments as $t_1(\eps)$ and $t_2(\eps)$, giving }
\beq
t_3(\eps) &=& \frac{\ts}{3\alpha^2} \left[ \left( \frac{E_0}{3\eps} \right)^{3/2} -1 \right] \label{eq:t11} \eqc \\
t_4(\eps) &=& \frac{\ts}{3\alpha^2} \left[ \left( 1+\frac{E_0}{\eps} \right)^{3/2} -1 \right] \label{eq:t12} \eqd
\eeq
For $t_{3,4}(\eps)>\lambda_0$ we find
\beq
\eps< \eps_3 = \frac{E_0}{3\left( 1+\frac{3\alpha^2\lambda_0}{\ts} \right)^{2/3}} \label{eq:eps11} \eqc \\
\eps< \eps_4 = \frac{E_0}{\left( 1+\frac{3\alpha^2\lambda_0}{\ts} \right)^{2/3} -1} \label{eq:eps21} \eqc 
\eeq
while for $t_{3,4}(\eps)>t_c$ we obtain
\beq
\eps< \eps_{c3} = \frac{E_0}{3\alpha^2} \label{eq:epsx1} \eqc \\
\eps< \eps_{c4} = \frac{E_0}{\alpha^2 -1} \label{eq:epsx2} \eqc
\eeq
respectively.

With these definitions, we sum up the results of the analytical calculation. 

We begin with the case $t_c<\lambda_0$:
\beq
L(t,\eps<\eps_3<E_0/3) = 3I_0 \epseb{1/3} \frac{\left( \frac{t}{\lambda_0} \right)^2}{1+3\left( \frac{t}{\lambda_0} \right)^2} \nonumber \\
\times \left( 1+\frac{t}{\ts} \right)^{2/3} e^{-\epse \left( \alpha_g+\frac{t}{\ts} \right)^2} \label{eq:I11} \eqc 
\eeq
\beq
L(t,\eps_3<\eps<\eps_{c3}<E_0/3) = 3I_0 \epseb{1/3} \frac{\left( \frac{t}{\lambda_0} \right)^2}{1+3\left( \frac{t}{\lambda_0} \right)^2} \nonumber \\
\times \left( 1+\frac{t}{\ts} \right)^{2/3} e^{-\epse \left( \alpha_g+\frac{t}{\ts} \right)^2} \label{eq:I12} \eqc 
\eeq
\beq
L(t,\eps_{c3}<\eps<E_0/3) = 3I_0 \epseb{1/3} \left( \frac{t}{\lambda_0} \right)^2 \nonumber \\
\times \left( 1+\frac{t}{\ts} \right)^{2/3} e^{-\epse \left( \alpha_g+\frac{t}{\ts} \right)^2} \label{eq:I13} \eqc 
\eeq
\beq
L(t,E_0/3<\eps) = 3I_0 \epseb{1/3} e^{-\epse} \frac{\left( \frac{t}{\lambda_0} \right)^2}{1+\frac{t}{2t_4}} \left[ 1-\frac{t}{\lambda_0} \right] \label{eq:I14} \eqd
\eeq
For $t_c>\lambda_0$ the analytical calculation yields for $\eps<E_0/3$
\beq
L(t<t_c,\eps<E_0/3) = 3I_0 \epseb{1/3} \frac{\left( \frac{t}{\lambda_0} \right)^2}{1+3\left( \frac{t}{\lambda_0} \right)^2} \nonumber \\
\times \left( 1+\frac{t}{\ts} \right)^{2/9} e^{-\epse \left( 1+\frac{t}{\ts} \right)^{2/3}} \label{eq:I21a}
\eeq
\beq
L(t>t_c,\eps<E_0/3) = I_0 \epseb{1/3} \nonumber \\
\times \left( \alpha_g+ \frac{t}{\ts} \right)^{2/3} e^{-\epse \left( \alpha_g+\frac{t}{\ts} \right)^2} \label{eq:I21b} \eqc
\eeq
which is the only case where { $L$ must be divided}. For $\eps>E_0/3$ we obtain
\beq
L(t,E_0/3<\eps<\eps_4) = 3I_0 \epseb{1/3} \frac{\left( \frac{t}{\lambda_0} \right)^2}{1+3\left( \frac{t}{\lambda_0} \right)^2} \nonumber \\
\times \left( 1+\frac{t}{\ts} \right)^{2/9} e^{-\epse \left( 1+\frac{t}{\ts} \right)^{2/3}} \label{eq:I22} \eqc 
\eeq
\beq
L(t,E_0/3<\eps_4<\eps) = 3I_0 \epseb{1/3} e^{-\epse} \frac{\left( \frac{t}{\lambda_0} \right)^2}{1+\frac{t}{2t_4}} \left[ 1-\frac{t}{\lambda_0} \right] \label{eq:I23} \eqd
\eeq
\begin{figure*}[t]
\begin{minipage}[t]{0.49\linewidth}
\centering \resizebox{\hsize}{!}
{\includegraphics{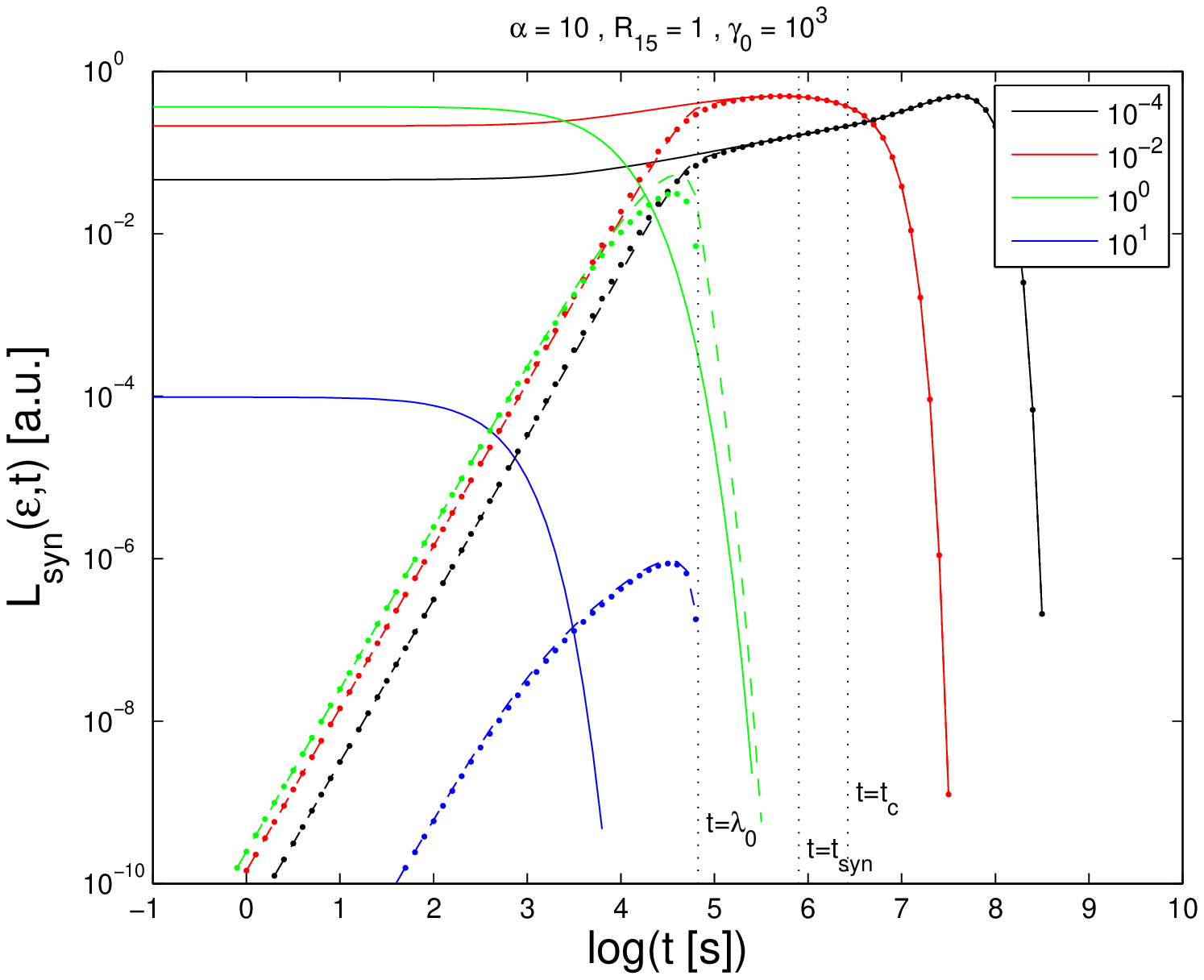}}
\end{minipage}
\hspace{\fill}
\begin{minipage}[t]{0.49\linewidth}
\centering \resizebox{\hsize}{!}
{\includegraphics{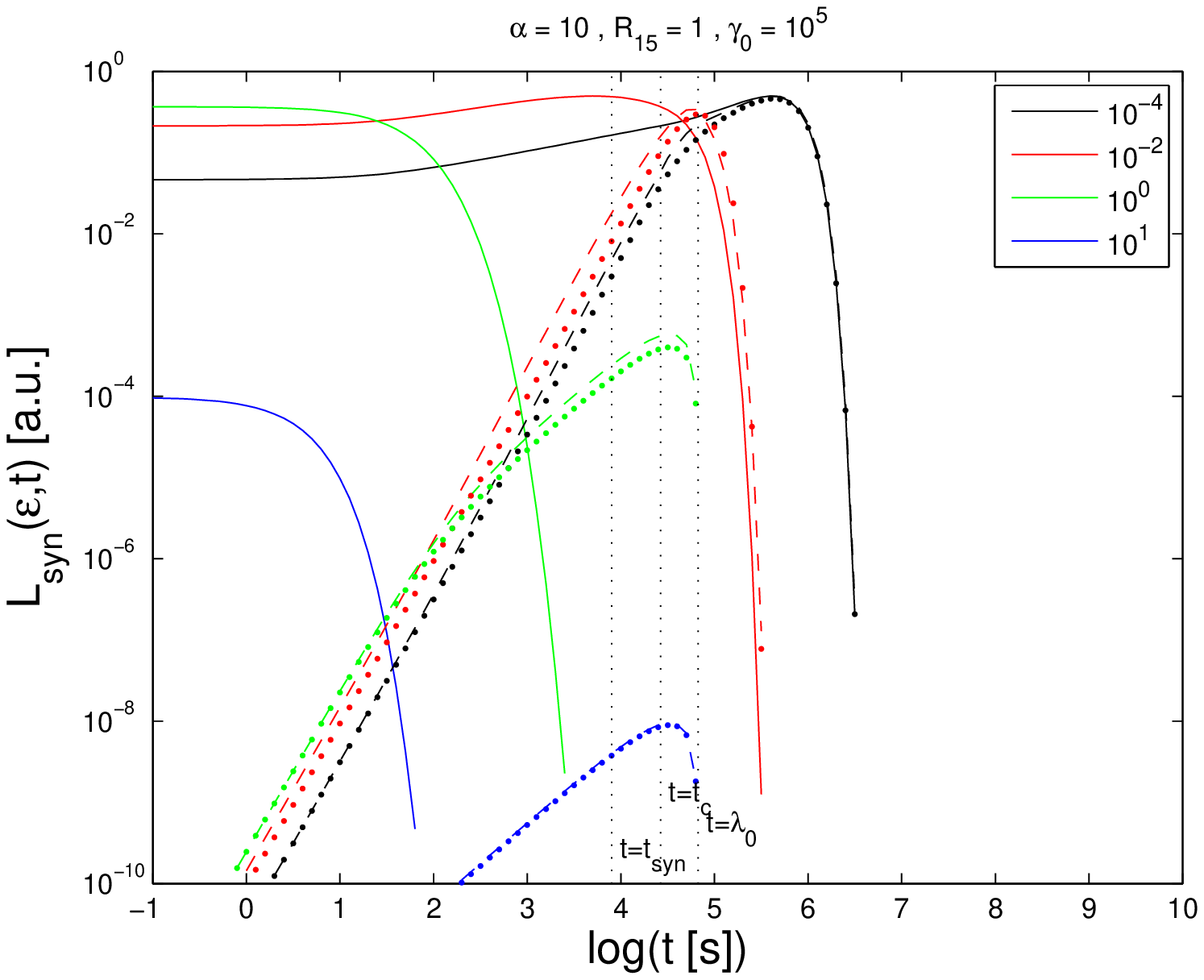}}
\end{minipage}
\caption{Unretarded (full), as well as numerical (dashed) and analytical (dotted) retarded lightcurve for $\alpha\ll 1$ and two cases of $\gamma_0$ { over a logarithmic time-axis}. The values of $\eps$ in the legend are given in units of $E_0$. The curves are normalized with $I_0$ and we set $b=1$.}
\label{fig:a10}
\end{figure*} 

In Figure \ref{fig:a10} we compare the analytical results with the numerical results, and achieve quite good agreement. The unretarded lightcurve is shown again for comparison.

Since the basic properties of the plot are the same as in Figure \ref{fig:a01}, we do not need to repeat them here. The problem with $t_4$ and $\eps_4$, mentioned in the discussion for Figure \ref{fig:a01}, is evident here, again.

%
%
\section{Discussion} \label{sec:dis}
\begin{figure*}[t]
\begin{minipage}[t]{0.49\linewidth}
\centering \resizebox{\hsize}{!}
{\includegraphics{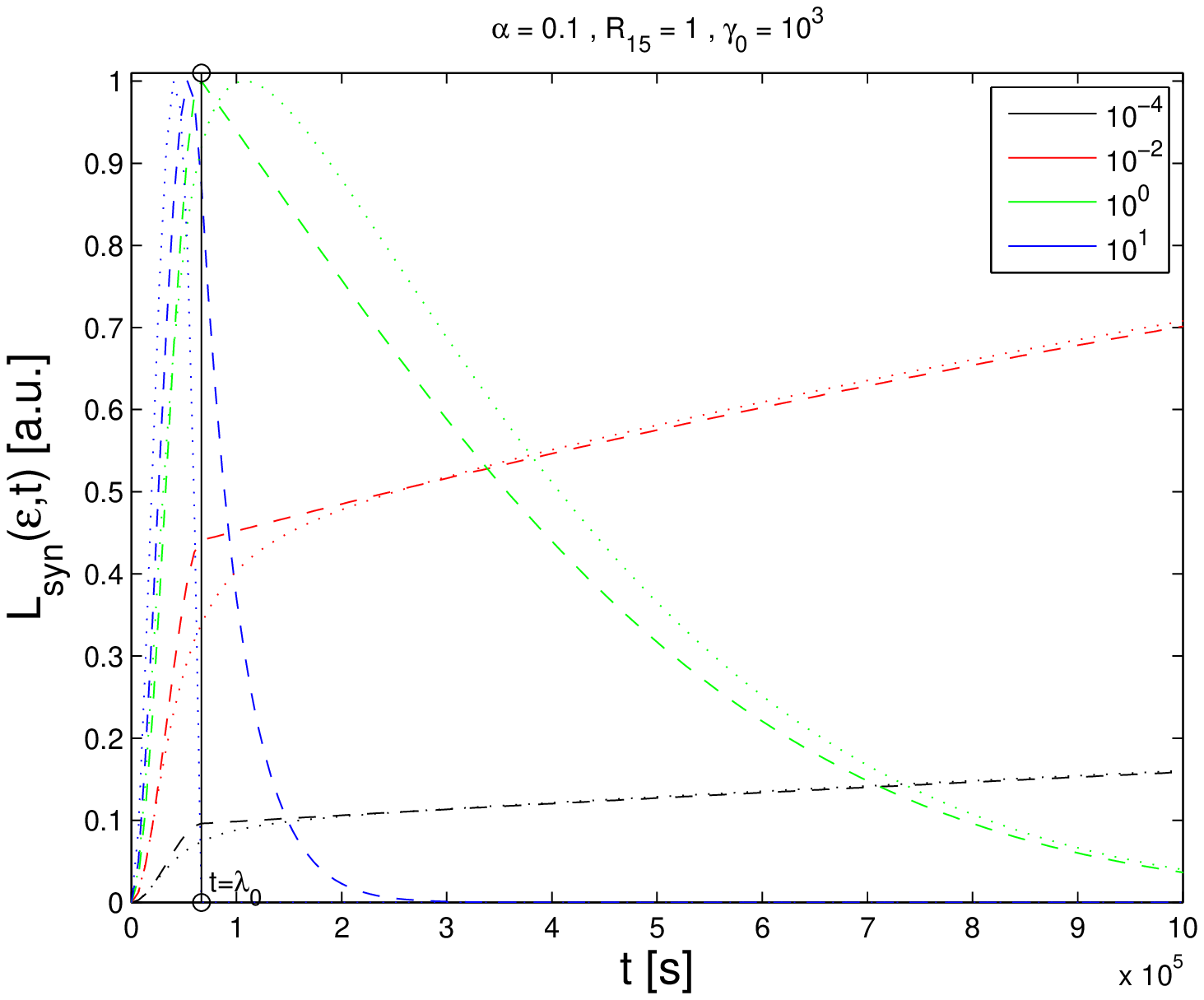}}
\end{minipage}
\hspace{\fill}
\begin{minipage}[t]{0.49\linewidth}
\centering \resizebox{\hsize}{!}
{\includegraphics{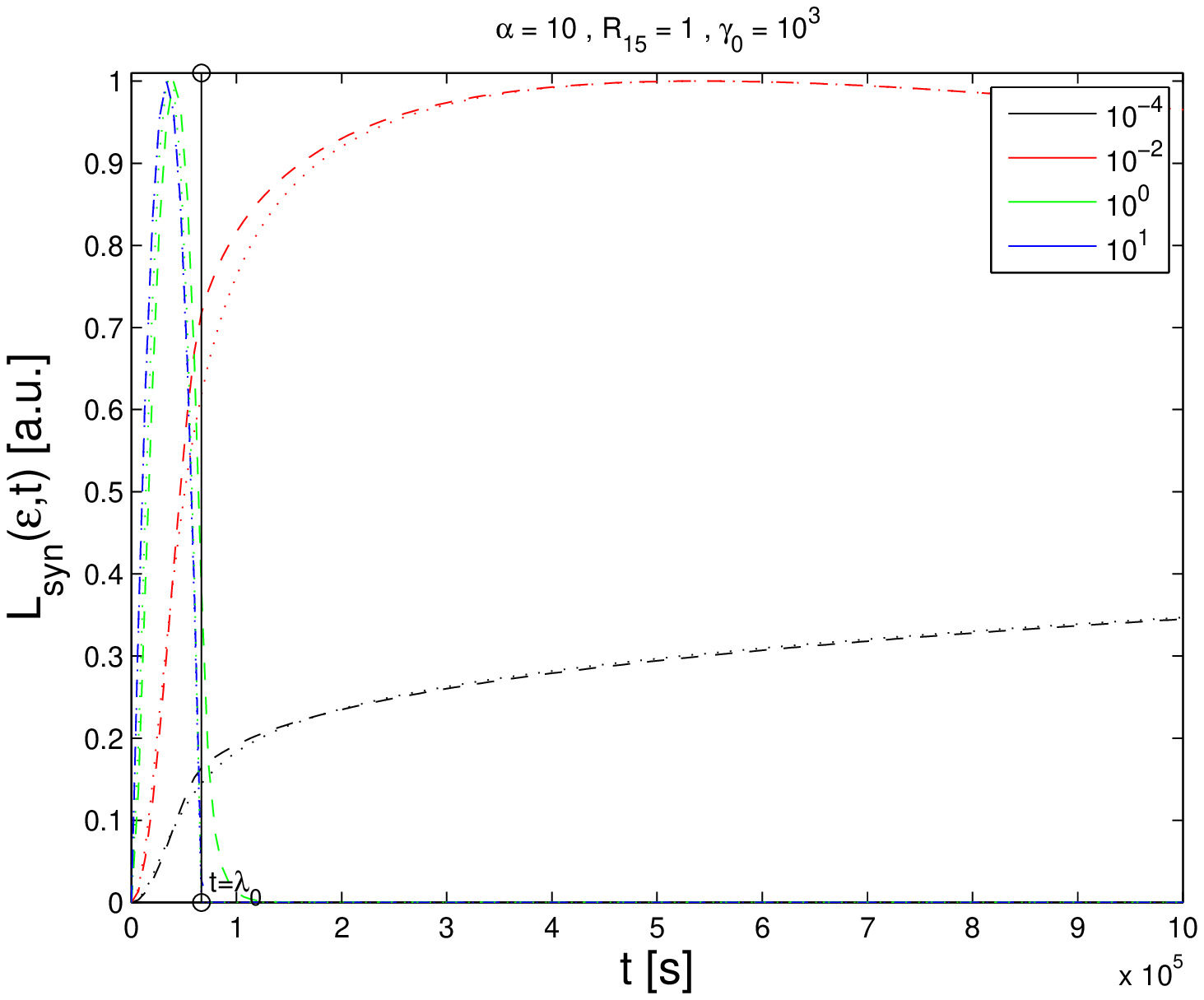}}
\end{minipage}
\newline
\begin{minipage}[t]{0.49\linewidth}
\centering \resizebox{\hsize}{!}
{\includegraphics{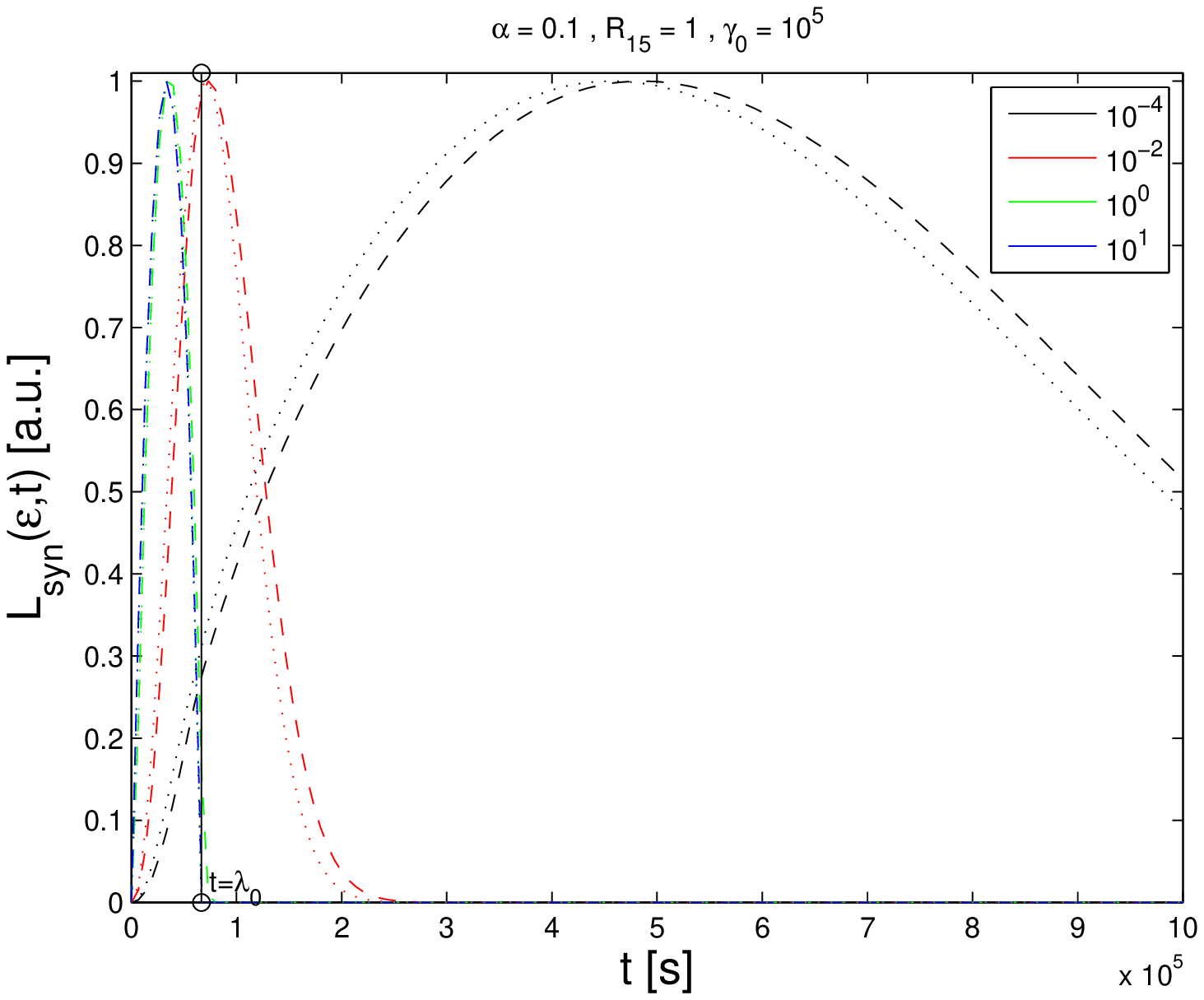}}
\end{minipage}
\hspace{\fill}
\begin{minipage}[t]{0.49\linewidth}
\centering \resizebox{\hsize}{!}
{\includegraphics{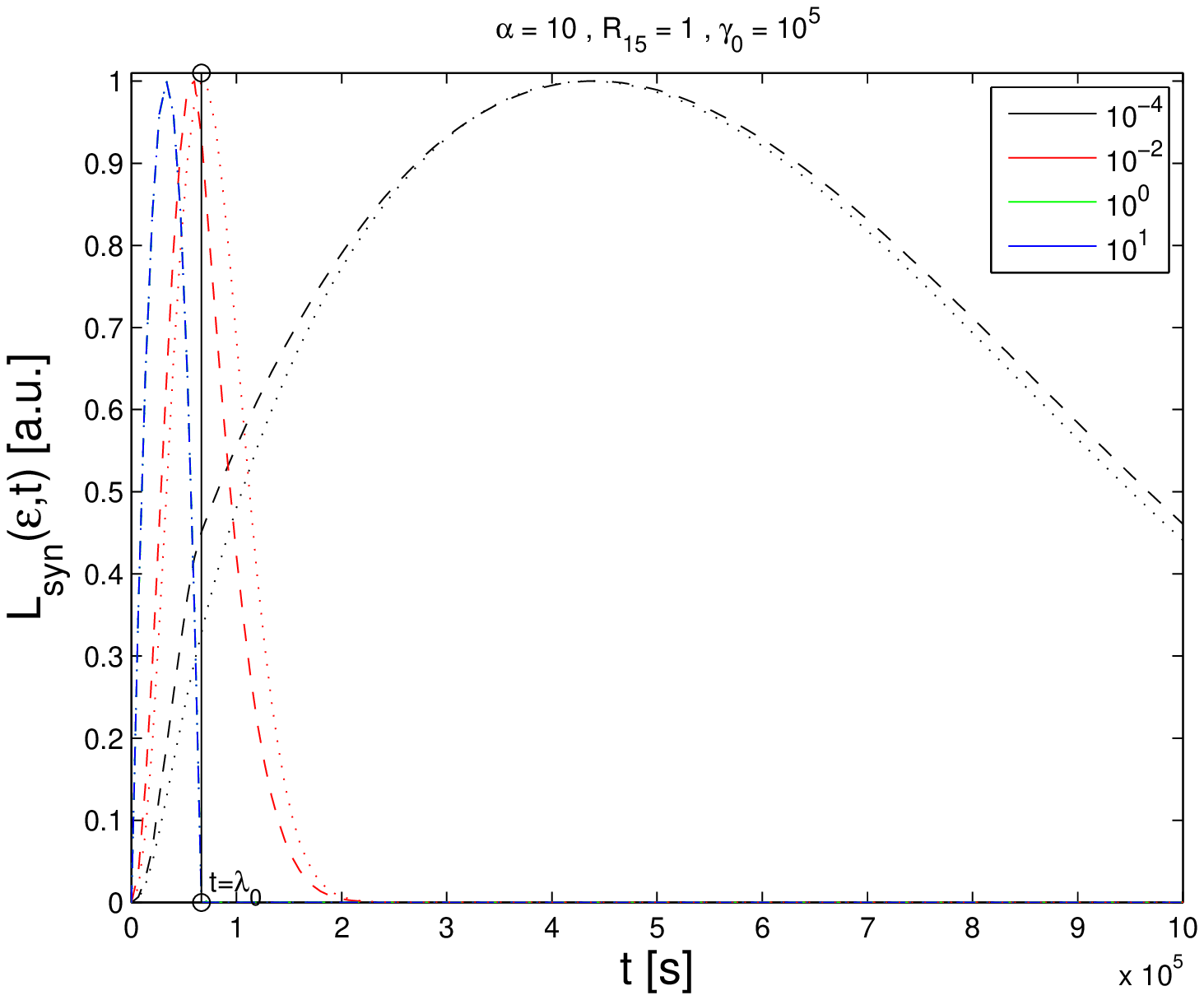}}
\end{minipage}
\caption{Analytical (dotted) and numerical (dashed) retarded lightcurves { over a liner time-axis}. The parameters are given at the top and values of $\eps$ in the legend are given in units of $E_0$. The curves are normalized by the respective maximum. The vertical line marks the light-crossing time scale $\lambda_0$, and the horizontal range is $15 \lambda_0$.}
\label{fig:linneu}
\end{figure*} 

{In Figures \ref{fig:a01} and \ref{fig:a10} we show lightcurves in a logarithmic plot, which has the advantage of having several cases in one plot. This makes it much easier to compare variability aspects which occur on very different time scales. Our discussion will focus on these logarithmic plots. On the other hand, lightcurves are commonly displayed in a linear plot, which highlights the behavior of lightcurves around their respective maxima. We present such linear plots in { Figure \ref{fig:linneu}. The results are completely compatible.} }

Comparing Figures \ref{fig:a01} and \ref{fig:a10} one can see that there are some points, where the results are similar, and some other points, which are remarkably different. 

First of all, {we note that the ``variability time scale'' of any given lightcurve is determined by its { global} maximum. Thus,} the minimal variability time scale, { which is possible at all,} is given by the light-crossing time scale,{ since the} source is evenly contributing to the radiative output. If the source only partially radiates, the variability time scale can be much lower (Eichmann et al. 2010). {The rising phase until $\lambda_0$ is dominated by the source geometry, giving a $t^2$-dependence up to the break times $t_2(\eps)$ for $\alpha\ll 1$ and $t_4(\eps)$ for $\alpha\gg 1$, respectively. If $t_{2,4}(\eps)<\lambda_0$, the lightcurve exhibits a break to a $t^1$-dependence. Otherwise the spectrum breaks directly to the unretarded lightcurve at $\lambda_0$.\footnote{The powers of $t$ depend sensitively on the chosen geometry. E.g. for a cylindrical source the power is reduced by unity giving a $t^1$- and a flat $t^0$-dependence.}}

{ Secondly, for larger initial electron energies $\gamma_0$ the variability time scale is much reduced compared to lower initial electron energies. Hence, in the low-energetic frequency bands the variability time scale shifts closer to $\lambda_0$ for larger $\gamma_0$. Thus, one can get information about the initial electron energy by observing the peak times of different frequency bands.}

The plots for the high-energetic cases ($\gamma_0=10^5$) look quite similar in both cases of $\alpha$, { since} $t_c$ is smaller than $\lambda_0$, and the lightcurves are the same for $t>t_c$. However, they can be distinguished by the high-energetic frequency bands. { Both are less luminous for $\alpha\gg 1$ compared to $\alpha\ll 1$, because} the synchrotron SED exhibits a broken power-law for $\alpha\gg 1$, leading to a decreased flux for high energies compared to the $\alpha\ll 1$ case (cf. SBM). Additionally, the break in the lightcurve from the quadratic {time-}dependence to the linear {time-}dependence takes place a factor $3\alpha^2$ earlier in the $\alpha\gg 1$ case than for $\alpha\ll 1$, { since} the unretarded lightcurve cuts off much earlier for $\alpha\gg 1$ than for $\alpha\ll 1$. Thus, the sum over all unretarded lightcurves of each slice (that is the retarded lightcurve) for $\alpha\gg 1$ must be less luminous and increase less strongly than for $\alpha\ll 1$. 

The low-energetic cases ($\gamma_0=10^3$) { differ strongly for $\alpha$ larger and smaller than unity}, since $t_c$ is larger than $\lambda_0$. The features, which are plainly visible in the unretarded lightcurve, are thus also visible in the retarded lightcurve. The lightcurve for $\eps=10E_0$ cuts off in both cases at around $\lambda_0$. { However}, for $\alpha\gg 1$ the break is clearly visible in the lightcurve, which is due to the faster cooling of the electrons in this case of $\alpha$. Similarly, the lightcurve for $\eps=E_0$ cuts off for $\alpha\gg 1$ at $\lambda_0$ because of the faster cooling, and for $\alpha\ll 1$ the lightcurve shows a rather broad exponential decay.\footnote{{ ``Broad'' and ``narrow'' are related to the appearance in the logarithmic plot. In a linear plot these features might look differently, since the widths in the logarithmic plot must be related to the order of magnitude examined, of course.}} The lightcurves for $\eps=10^{-2}E_0$ can be distinguished quite well, since the lightcurve for $\alpha\ll 1$ shows a rapid increase to the narrow maximum, while the light curve for $\alpha\gg 1$ exhibits a broad and flat maximum, which covers almost two orders of magnitude in time. {As stated above, because of the broken power-law in the SED for $\alpha\gg 1$ the maximum of the SED is attained at $\eps=E_0/\alpha^2$, which is in our example the lightcurve for $\eps=10^{-2}E_0$. This explains the broad maximum, and also why the higher energies are again less luminous compared to the $\alpha\ll 1$ case.} In the lightcurves for $\eps=10^{-4}E_0$ there is only a very slight difference, since the rising part after $\lambda_0$ sets in earlier for $\alpha\gg 1$ than for $\alpha\ll 1$. However, the increase is quite small until $t_c$ and detailed observations are needed to distinguish the models.

Obviously, for smaller emission regions the light-crossing time scale $\lambda_0$ is reduced and the effects of the time-dependent cooling will be even more pronounced {with very short variability time scales}. Additionally, a smaller emission region leads to a larger $\alpha$ according to equation (\ref{eq:alphascaled}). Thus, the time-dependent SSC cooling is quite important for models which assume a small emission region to explain the rapid variability in blazars, {like those cited in the introduction}. 

{ We numerically checked if our results are also valid for different injection scenarios, such as a power-law, and obtained qualitatively similar results as one can see in { Figure \ref{fig:PLplots} of} appendix \ref{app:plplot}. Thus, we are confident that the discussion presented here is robust.}


{
\subsection{Caveats of the approach} \label{sec:dis_cav}

Of course, our approach is simplified, leaving aside several, possibly important aspects.

As stated in section \ref{sec:geo}, we assume that the source is spatially homogeneous. Additionally, we neglect the ``internal'' retardation of the inverse Compton processes. The time, a photon travels before it is inverse Compton scattered, should on average be much less than $\lambda_0$. Thus, the effect on the light curve is small, apart from a short delay of the SSC light curve (which we do not calculate in this paper). However, cooling might be affected, since the SSC cooling should be similarly delayed as the SSC light curve. On the other hand, as shown in Schlickeiser (2009) and Zacharias \& Schlickeiser (2010), the SSC cooling is orders of magnitude quicker than the synchrotron and external Compton cooling, which should more than compensate the small retardation delay.

We do not expect a delay effect for the external Compton scattering, since they are present all the time and are more or less homogeneously distributed (ZSb, Sokolov \& Marscher 2005).

We also do not discuss the acceleration of the electrons. This might have a significant impact on the light curve, since the variability is governed by the longest time scale (in our case just $\lambda_0$ and $t_{syn}$, with an influence by $t_c$ for $\alpha\gg 1$). If the acceleration takes particularly long the effects due to retardation or cooling would be washed out. If the acceleration takes only a small amount of time, it will certainly influence the rising phase of the light curve, but its effect on the main part of the variability will not be significant. This argument might not hold for very small emission regions. 

On the other hand, the acceleration time scale is only important, if the acceleration and radiation zone are spatially and temporally coincident. This is still under debate, and only recently these conditions are incorporated in numerical studies (e.g. Weidinger et al. 2010; Weidinger \& Spanier 2010). Thus, we follow the assumption or simplification that acceleration and radiation are (especially) temporally separated. Then, the acceleration time scale does not influence the resulting light curve.

Chiaberge \& Ghisellini (1999), and also e.g. Kataoka et al. (2000) or Li \& Kusunose (2000), used similar assumptions in their numerical analysis. In fact, most theoretical investigations use numerical schemes with the advantage of employing more and more realistic scenarios, such as time-dependency (B\"ottcher \& Chiang, 2002), inclusion of shock acceleration (Sokolov et al., 2004), hydrodynamic simulations (Mimica et al., 2004; Cabrera et al., 2013), and multizone models that incorporate the full retardation of all processes (Graff et al., 2008, Joshi \& B\"ottcher 2011). In our analytical discussion we are for obvious reason not able to include all these details. That is why we focus on the details of the time-dependency, showing analytically that time-dependent effects, especially from SSC, are very important for rapid flares.

}

%
%
\section{Conclusions} \label{sec:con}

In this paper we {introduced our approach to} calculate theoretical synchrotron lightcurves for flaring blazars, where the { radiating relativistic} electrons are cooled by the combined synchrotron, external Compton and time-dependent SSC mechanisms. This complements the recent series of papers (Schlickeiser 2009; Schlickeiser et al. 2010 (SBM); Zacharias \& Schlickeiser 2010, 2012a (ZSa), 2012b (ZSb)) on the effects of the combined cooling on the SED. Lightcurves show the intensity of a specific frequency band over time. Thus, they are a perfect tool to analyze the flaring behavior of blazars in different energies, such as correlations between different frequency bands.

We were able to show that the { synchrotron} lightcurves exhibit a different form, if the time-dependent nature of the SSC cooling is taken into account, compared to the usual time-independent approaches. For that we first derived a formula to calculate the lightcurve from the intensity distribution, where we introduced the retardation due to the finite size of the radiation source. Using the intensities derived by SBM for the time-independent and the time-dependent cooling scenarios, we calculated the resulting lightcurves.

{{ Our calculations highlight the differences between the usual linear and the time-dependent cooling scenarios, giving us} confidence that the important effects in the lightcurves are really due to the different cooling terms, and are not hidden by other effects. 

The main results can be summarized as follows.

(1) Until the light crossing time scale $\lambda_0$ is reached, the initial { synchrotron} lightcurves depend strongly on the geometry of the source. In our example of a spherical source the lightcurve increases $\propto t^2$, which, depending on { the synchrotron photon energy} $\eps$, is followed by a linear $t$-dependence. We note, however, that this rising phase might be hard to observe depending on the sampling rate in specific frequency bands. 

(2) If { the transition time $t_c$ from time-dependent SSC to linear cooling is larger than the light crossing time scale (i.e., $t_c>\lambda_0$), the effects of the rapid SSC cooling are clearly visible for $t>\lambda_0$.} The lightcurves exhibit their respective maximum up to an order of magnitude earlier, { if the electrons cool initially by the time-dependent SSC process.} In this cooling regime variability can be 10 times faster than in the linear cooling regime. The different { spectral} powers below and above { the transition time} $t_c$ probably need very precise measurements to be distinguishable in the data of blazars.

(3) The lightcurves are rather similar, if $t_c<\lambda_0$, since the effects of { the time-dependent} SSC cooling are { smeared} out by the retardation.

The results (2) and (3) obviously depend sensitively on the source parameters. In very compact emission regions with a short light crossing time scale and { a large injection parameter} $\alpha$, the effects of the time-dependent SSC cooling are most significant.} This in combination with the strong effects on the SED (e.g. ZSb) should help to clearly discriminate between different models, and to restrict the parameter space. { In this context} theoretical prediction of the SSC and EC lightcurve are also mandatory, and we intend to publish the results in a future work, {where also a much deeper discussion of correlations is possible}.

To conclude, we argue for a wide utilization of the time-dependent SSC cooling scenario, at least for the modeling of rapid flares in blazars, {where compact emission regions are necessary.}

%
%
\acknowledgements

We thank the anonymous referee for constructive comments, which helped significantly to improve the manuscript. \\
We acknowledge support from the German Ministry for Education and Research (BMBF) through Verbundforschung Astroteilchenphysik grant 05A11PC1 and the Deutsche Forschungsgemeinschaft through grant Schl 201/23-1. 

%
%

%
%
\appendix
\section{Calculation of the intermediate part} \label{app:intpart}

The intermediate time regime $t_{1,2}<t<\lambda_0$ requires another approach. The approximations of the other regimes cannot be used here.

\beq
L(t_{1,2}<t<\lambda_0,\eps) &=& 6I_0 \epseb{1/3} \intl_0^{t/\lambda_0} \left( 1+\frac{t-\lambda_0\lambda}{\ts} \right)^{2/3} e^{-\epse \left( 1+\frac{t-\lambda_0\lambda}{\ts} \right)^{2}} \left( \lambda-\lambda^2 \right) \td{\lambda} \nonumber \\
&\approx& 6I_0 \epseb{1/3} \left( 1+\frac{t}{\ts} \right)^{2/3} e^{-\epse\left( 1+\frac{t}{\ts} \right)^2} \intl_0^{t/\lambda_0} \left( 1-\frac{2}{3}\frac{\lambda_0\lambda}{\ts+t} \right) e^{2\epse \left( 1+\frac{t}{\ts} \right) \frac{\lambda_0\lambda}{\ts}} \left( \lambda-\lambda^2 \right) \td{\lambda} \nonumber \\
&=& 6I_0 \epseb{1/3} \left( 1+\frac{t}{\ts} \right)^{2/3} e^{-\epse\left( 1+\frac{t}{\ts} \right)^2} \nonumber \\
&& \times \intl_0^{t/\lambda_0} \left( \lambda - \left( 1+\frac{2\lambda_0}{3(\ts+t)} \right)\lambda^2 + \frac{2\lambda_0}{3(\ts+t)}\lambda^3  \right) e^{\frac{2\eps\lambda_0}{E_0\ts} \left( 1+\frac{t}{\ts} \right) \lambda} \td{\lambda} 
\eeq
Integrating by parts and approximating to first order yields
\beq 
L(t_{1,2}<t<\lambda_0,\eps) \approx 3I_0 \epseb{-2/3} e^{-\epse} \frac{\ts^2}{\lambda_0^2} \left( \frac{t}{\ts} \right) \left[ 1-\frac{t}{\lambda_0} \right] \eqc
\eeq
where we also approximated for $t<\ts$. This is the result (\ref{eq:I03}), which fits very well the numerical solution for intermediate times.

\section{Power-law plots} \label{app:plplot}

\begin{figure*}[t]
\begin{minipage}[t]{0.49\linewidth}
\centering \resizebox{\hsize}{!}
{\includegraphics{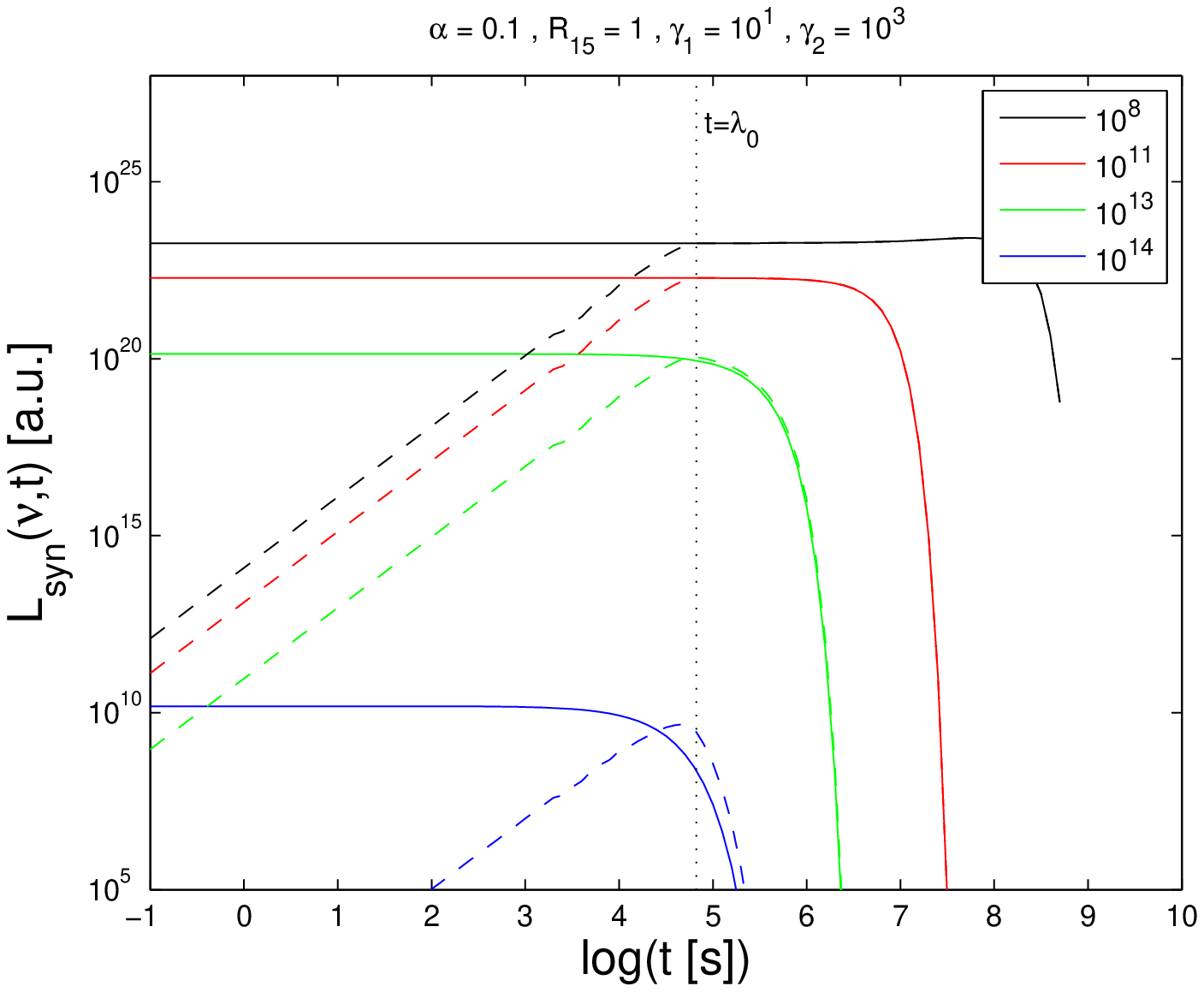}}
\end{minipage}
\hspace{\fill}
\begin{minipage}[t]{0.49\linewidth}
\centering \resizebox{\hsize}{!}
{\includegraphics{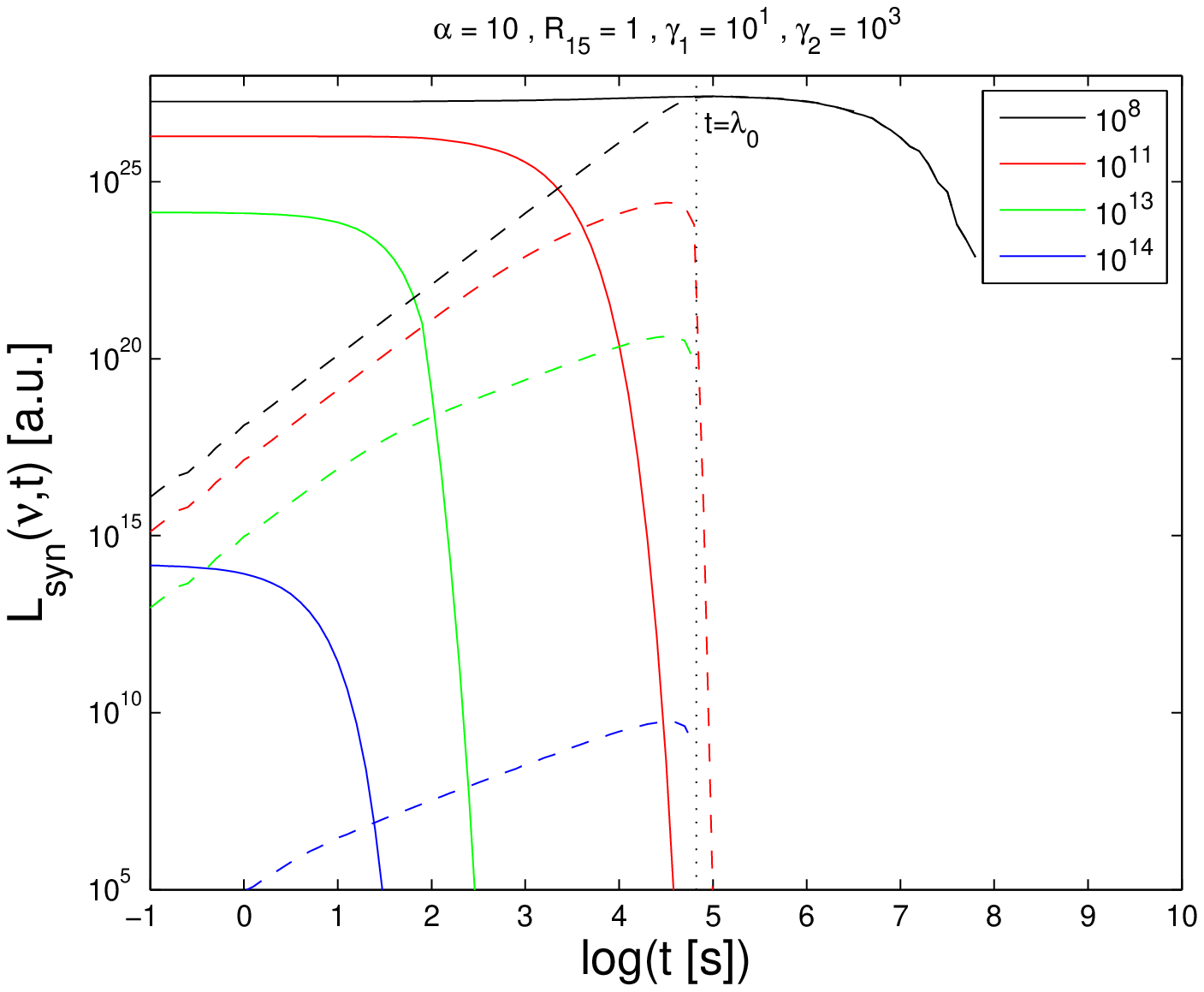}}
\end{minipage}
\newline
\begin{minipage}[t]{0.49\linewidth}
\centering \resizebox{\hsize}{!}
{\includegraphics{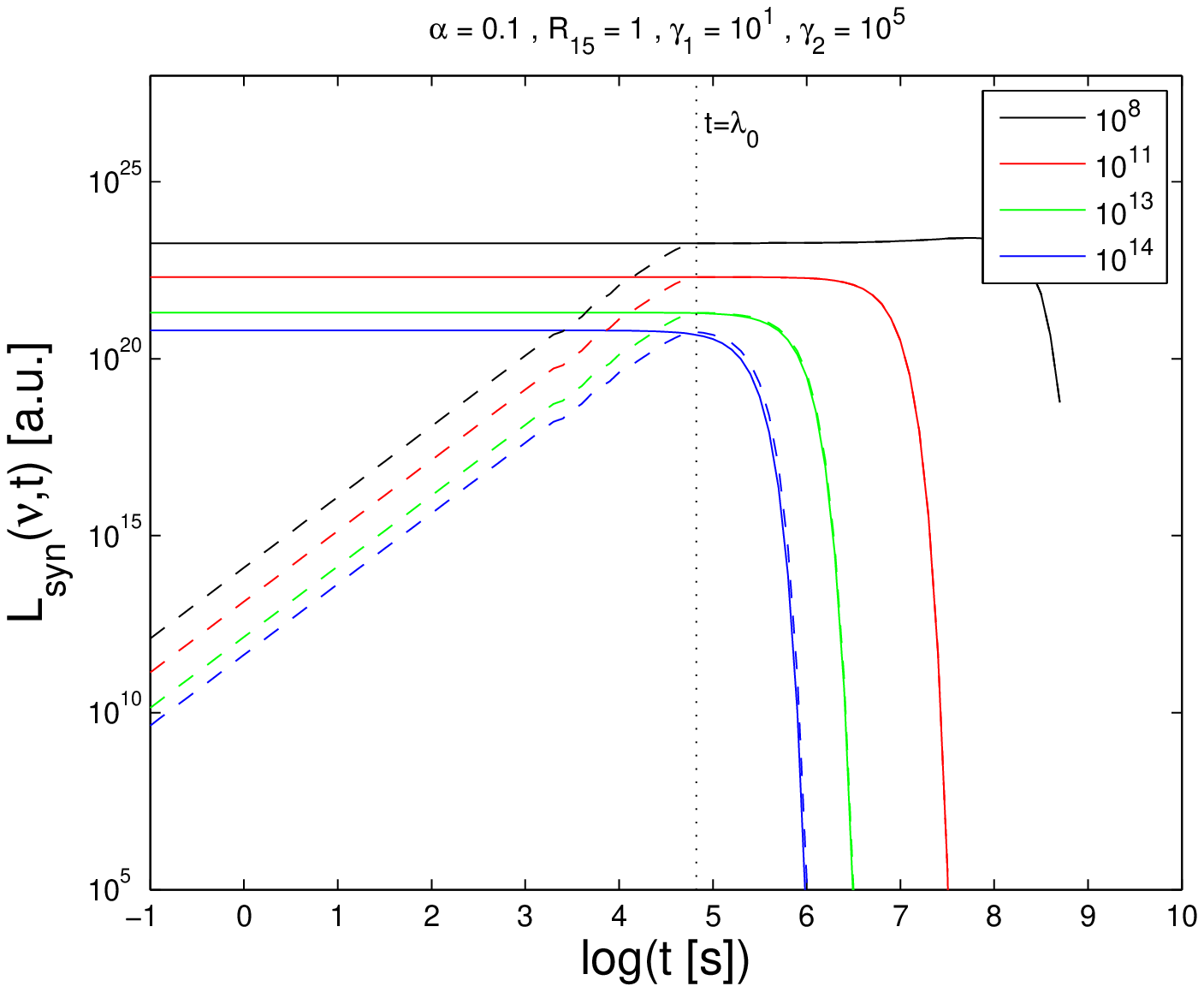}}
\end{minipage}
\hspace{\fill}
\begin{minipage}[t]{0.49\linewidth}
\centering \resizebox{\hsize}{!}
{\includegraphics{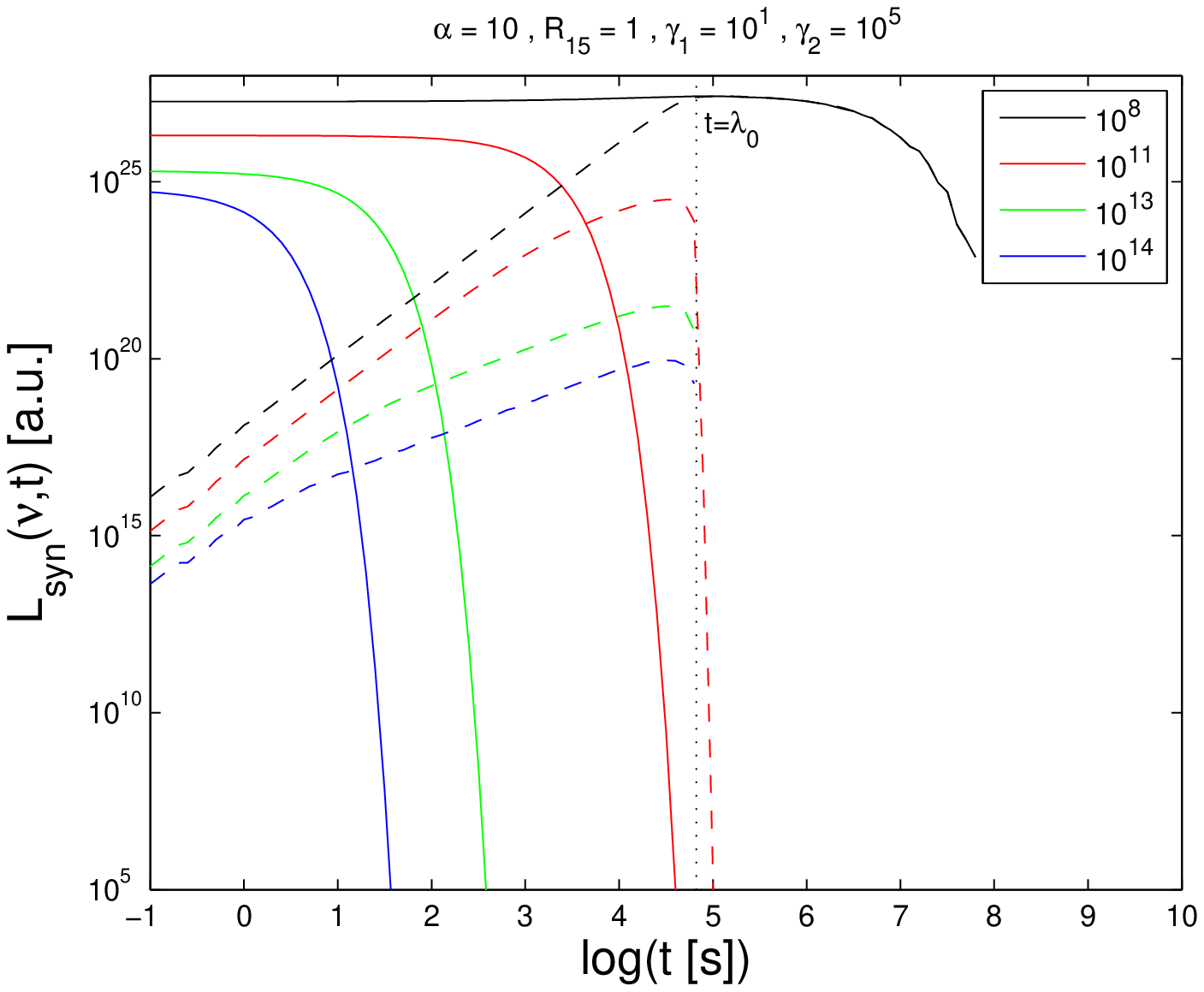}}
\end{minipage}
\caption{Unretarded (full) and numerical (dashed) retarded lightcurves for a power-law injection with spectral index $s=2$. The parameters are given at the top. The legend is for frequencies $\nu$. The vertical dotted line marks the light-crossing time scale $\lambda_0$.}
\label{fig:PLplots}
\end{figure*} 

In order to check if our analytical results can be regarded as qualitatively general, we performed a numerical integration of equation (\ref{eq:L}) with a power-law injection of the form
\begin{eqnarray}
S(\gamma,t_{em}) = q_0 \gamma^{-s} \HF{\gamma-\gamma_1} \HF{\gamma_2-\gamma} \DF{t_{em}} \label{eq:PLinj} \eqd
\end{eqnarray} 
The differential equation (\ref{eq:kineq}) with this type of injection has been solved by Zacharias \& Schlickeiser (2010), and we use their result in equation (\ref{eq:monoint}) in order to calculate the intensity distribution.

For the illustrative case $s=2$ the results are plotted in Fig. \ref{fig:PLplots} for two cases of $\alpha$ and two cases of the upper limit $\gamma_2$, respectively. Without going into details, one can see that the results discussed in section \ref{sec:dis} are qualitatively recovered, which gives us confidence that our approach is robust.

\end{document}